\documentclass[journal,comsoc]{IEEEtran}
\usepackage{indentfirst}
\usepackage{setspace}
\usepackage[nospace,compress,sort]{cite}
\usepackage{graphicx}
\usepackage{amstext}
\usepackage{algorithm}
\usepackage{algorithmic}
\usepackage{mathrsfs}
\usepackage{amssymb}
\usepackage{amsmath}
\usepackage{epstopdf}
\usepackage{CJK}
\usepackage{multicol}
\usepackage{stfloats}
\usepackage{url}
\usepackage{color}
\usepackage{enumerate}
\allowdisplaybreaks[4]

\newtheorem{Theorem}{Theorem}

\newtheorem{Remark}{Remark}

\ifCLASSINFOpdf
\else
\fi
\begin{document}

\title{\huge{The Distributed MIMO Scenario: Can Ideal ADCs Be Replaced by Low-resolution ADCs?}}


\author{Jide Yuan,~\IEEEmembership{Student Member,~IEEE},
Shi Jin,~\IEEEmembership{Member,~IEEE},\\
Chao-Kai Wen,~\IEEEmembership{Member,~IEEE}, and Kai-Kit Wong,~\IEEEmembership{Fellow,~IEEE}
\thanks{J. Yuan and S. Jin (corresponding author) are with the National Mobile Communications Research Laboratory, Southeast University, Nanjing 210096, P. R. China (e-mail: $\rm \{yuanjide,jinshi\}@seu.edu.cn$).}
\thanks{C. Wen is with the Institute of Communications Engineering, National Sun Yat-sen University, Kaohsiung 804, Taiwan (e-mail: $\rm ckwen@ieee.org$).}
\thanks{K. K. Wong is with the Department of Electronic and Electrical Engineering, University College London, London WC1E 7JE, United Kingdom (e-mail: $\rm kai\text{-}kit.wong@ucl.ac.uk$).}
\thanks{This work was supported in part by the National Science Foundation (NSFC) for Distinguished Young Scholars of China with Grant 61625106 and the National Natural Science Foundation of China under Grant 61531011.}
}

\maketitle
\begin{spacing}{0.9}
\begin{abstract}
This letter considers the architecture of distributed antenna system, which is made up of a massive number of single-antenna remote radio heads (RRHs), some with full-resolution but others with low-resolution analog-to-digital converter (ADC) receivers. This architecture is greatly motivated by its high energy efficiency and low-cost implementation. We derive the worst-case uplink spectral efficiency (SE) of the system assuming a frequency-flat channel and maximum-ratio combining (MRC), and reveal that the SE increases as the number of quantization bits for the low-resolution ADCs increases, and the SE converges as the number of RRHs with low-resolution ADCs grows. Our results furthermore demonstrate that a great improvement can be obtained by adding a majority of RRHs with low-resolution ADC receivers,
if sufficient quantization precision and an acceptable proportion of high-to-low resolution RRHs are used.
\end{abstract}

\begin{IEEEkeywords}
Distributed antenna system, low-resolution ADC receiver, massive MIMO, spectral efficiency.
\end{IEEEkeywords}

%
\IEEEpeerreviewmaketitle

\section{Introduction}
\IEEEPARstart{M}{assive} multiple-input multiple-output (MIMO) antennas are widely anticipated as one key 5G technology \cite{6736746,7395265}. A distributed setup, with antennas spread over a large area connecting to a baseband unit (BBU) via a fronthaul network, has considerable advantages in both spectral efficiency (SE) and energy efficiency (EE) due to reduced distances between users and the remote radio heads (RRHs) as well as the rich diversity against shadow fading \cite{5770665,123456789}. In \cite{5770665}, it was revealed that enormous gains can be achieved by distributed massive MIMO antennas if compared to the centralized counterpart. A recent work \cite{123456789} presented the cell-free massive MIMO system, with all RRHs serving all users, outperforming the traditional centralized massive MIMO in terms of throughput.

Nevertheless, the hardware expenditure and power consumption grows with the number of antennas. This has motivated the adoption of low-resolution analog-to-digital converters (ADCs) in antennas to save cost \cite{7307134,MollenCLH16a}. In\cite{MollenCLH16a}, the SE for a multi-user massive MIMO with arbitrary quantization was studied. The result can be applied into frequency-selective MIMO channel, and demonstrated that the antenna with 3-bits quantization can unleash most of the performance gains that the setup with all full-resolution ADC has. Mixed-ADC architectures have since been proposed, e.g., \cite{7562390,7579180}, where only a fraction of the antennas are equipped with full-resolution ADCs, with the rest low-resolution ADCs. In \cite{7562390}, a mixed-ADC framework for massive MIMO uplink was established by developing a family of detectors via probabilistic Bayesian inference. The results showed that using just a small number of full-resolution ADCs can achieve a significant gain.
\begin{figure}[t]
\centering
\includegraphics[width=8.5cm]{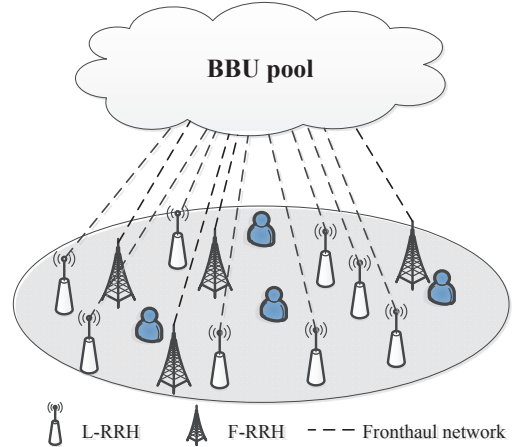}
\caption{The diagram of the distributed massive MIMO with mixed-ADC receivers, where some of the RRHs are equipped with full-resolution ADCs and others with low-resolution ADCs. All RRHs belong to one BBU.}
\end{figure}

In this letter, the focus is on the distributed massive MIMO setup, where RRHs with mixed ADC receivers are deployed. In particular, the RRHs with full-resolution ADCs (F-RRHs) or low-resolution ADCs (L-RRHs) are randomly distributed in the area. Our aim is to analyze the uplink SE performance assuming perfect knowledge of channel statistics and the use of a maximal-ratio combining (MRC) receiver. The effects of the user transmit SNR and the number of quantization bits of the low-resolution ADCs on the achievable performance are examined using the derived lower bounds of SE.
Our results demonstrate that distributed massive MIMO with mixed-ADC receivers 
possesses much higher SE than the centralized counterpart, and can realize most of the performance gains at much lower cost than the configuration with all F-RRHs.





\section{System Model}
Consider a multi-user distributed massive MIMO uplink with $K$ users and $M$ RRHs (see Fig.~1), in which each RRH and user has a single antenna. All of the RRHs in the network serve all users simultaneously. We assume that only a fraction of RRHs (e.g., $M_\text{f}$) are F-RRHs, while the rest ($M_\text{l}=M-M_\text{f}$) are L-RRHs. The locations of RRHs and users are randomly generated, and all of the RRHs are connected to a BBU.

We model the complex propagation coefficient between the $m$th RRH and the $k$th user by
\begin{equation}\label{propagation}
{g_{mk}} = {h_{mk}}\sqrt {{\beta _{mk}}},
\end{equation}
where $h_{mk}$ denotes the fast fading coefficient assumed to have zero-mean and unit-variance, and $\beta _{mk}$ is the large scale fading which can be factored into $\beta _{mk}=z_{mk}d_{mk}^{-\gamma}$, where $z_{mk}$ is a log-normal random variable with standard deviation $\sigma_{\text{shad}}$, $d_{mk}$ is the distance between the $m$th RRH and the $k$th user, and $\gamma$ is the path-loss exponent. We use $g_{\text{f},mk}$ and $g_{\text{l},mk}$ to specify the channel propagation from F-RRHs and L-RRHs, respectively.

In the uplink, the received signals of the $M_\text{f}$ F-RRHs can be packed into a vector
\begin{equation}\label{receive_full}
\setlength{\abovedisplayskip}{4pt}
\setlength{\belowdisplayskip}{4pt}
{\bf{y}_\text{f}} = \sqrt \rho {\bf{G}_\text{f}\bf{x}} + {\boldsymbol{\omega }_\text{f}},
\end{equation}
where $\bf{x}$ is the $K \times 1$ vector, whose $k$th element is the $k$th user symbol, $x_k$, with ${\mathsf E}[\left|x_k\right|^2]=1$. Also, ${\boldsymbol{\omega }_\text{f}\sim\mathcal{CN}\left(0,\bf{I}\right)}$ is the additive white Gaussian noise (AWGN), $\rho$ is the system SNR, and $\bf{G}_\text{f}$ is the {$M_\text{f} \times K$} channel matrix between the $M_\text{f}$ F-RRHs and the $K$ terminals with $[{{\bf{G}}_\text{f}}]_{mk}=g_{\text{f},mk}$. 
Assuming a frequency-flat channel, the quantization signal of $M_\text{l}$ L-RRHs with each a $B$-bit quantizer can be approximately depicted by the additive quantization noise model (AQNM) given by \cite{7308988}
\begin{equation}\label{receive_low}
{{{\bf{\tilde y}}}_{\rm{l}}} \approx \alpha {{\bf{y}}_{\rm{l}}} + {{\boldsymbol{\omega }}_{\rm{q}}} = \alpha \sqrt \rho  {{\bf{G}}_{\rm{l}}}{\bf{x}} + \alpha {\boldsymbol{\omega }_\text{l}} + {{\boldsymbol{\omega }}_{\rm{q}}},
\end{equation}
where ${\boldsymbol{\omega }_\text{l}\sim\mathcal{CN}\left(0,\bf{I}\right)}$,  ${{\boldsymbol{\omega }}_{\rm{q}}}$ is the quantization noise vector that is uncorrelated with ${\bf{\tilde y}}_\text{l}$. For the non-uniform scalar minimum mean squared error (MMSE) quantizer of a Gaussian random variable,
the values of $\alpha$, obtained by numerical computation, increase with $B$ and are upper bounded by 1 \cite{7307134}. In addition, $\bf{G}_\text{l}$ is the ${M_\text{l} \times K}$ propagation matrix between the $M_\text{l}$ L-RRHs and the $K$ terminals, whose $\left(i,j\right)$th entry is $[{{\bf{G}}_\text{l}}]_{mk}=g_{\text{l},mk}$.
Combining (\ref{receive_full}) and (\ref{receive_low}), the overall received signals can be modelled as
\begin{equation}\label{receive_all}
{\bf{y}} = \left[
{{{\bf{y}}^{T}_{\rm{f}}}} ~~{{{{\bf{\tilde y}}}^{T}_{\rm{l}}}}
\right]^{T}.
\end{equation}
Adopting an unbiased channel estimator at each RRH which can provide the exact mean of the statistic, we have
\begin{align}
{\mathsf E}\left[\hat{g}^{H}_{\text{f},{mk}}{g}_{\text{f},{mk}}\right]=\beta_{mk}~~\text{and}~~
{\mathsf E}\left[\hat{g}^{H}_{\text{l},{m'k}}{g}_{\text{l},{m'k}}\right]=\beta_{m'k},
\end{align}
where $\hat{g}_{\text{f},{mk}}$ and $\hat{g}_{\text{l},{m'k}}$ are the estimated channel coefficients from the $k$th user to the $m$th F-RRH and the $m'$th L-RRH, respectively.

In this letter, we consider that the MRC receiver is adopted
because of its significantly low complexity.
From (\ref{receive_all}), the received signal vector after MRC processing is given as
\begin{align}\label{MRC_all}
{\bf{r}}&\approx {\tilde{\bf{r}}} = \left[
{{\bf{\hat G}}_{\rm{f}}^H}~
{{\bf{\hat G}}_{\rm{l}}^H}
\right]{\bf{y}}.
\end{align}
From (\ref{MRC_all}), the $k$th element of $\tilde{\bf{r}}$ can be expressed as \begin{align}\label{MRC_k}
{\tilde{r}_k} &= \sqrt \rho  \left( {{\hat{\bf{g}}}_{{\rm{f}},k}^H{{\bf{g}}_{{\rm{f}},k}} + \alpha {\bf{\hat g}}_{{\rm{l}},k}^H{{\bf{g}}_{{\rm{l}},k}}} \right){x_k} \notag\\
&+ \sqrt \rho  \sum\nolimits_{i \ne k} {\left( {{\hat{\bf{g}}}_{{\rm{f}},k}^H{{\bf{g}}_{{\rm{f}},i}} + \alpha {\bf{\hat g}}_{{\rm{l}},k}^H{{\bf{g}}_{{\rm{l}},i}}} \right){x_i}} \notag\\
&+ \left( {{\hat{\bf{g}}}_{{\rm{f}},k}^H{{\boldsymbol{\omega }}_{\rm{f}}} + \alpha {\bf{\hat g}}_{{\rm{l}},k}^H{{\boldsymbol{\omega }}_{\rm{l}}}} \right) + {\bf{\hat g}}_{{\rm{l}},k}^H{{\boldsymbol{\omega }}_{\rm{q}}},
\end{align}
where ${\bf{g}}_{{\rm{f}},k}$, ${\bf{g}}_{{\rm{l}},k}$, ${\hat{\bf{g}}}_{{\rm{f}},k}$ and ${\hat{\bf{g}}}_{{\rm{l}},k}$ are the $k$th columns of ${\bf{G}}_{\rm{f}}$, ${\bf{G}}_{\rm{l}}$, ${\hat{\bf{G}}}_{\rm{f}}$ and ${\hat{\bf{G}}}_{\rm{l}}$, respectively.



\section{SE Analysis}
In this section, we derive the achievable SE of the network. Given that the unbiased channel estimators are employed at the RRHs, the received signal in (\ref{MRC_k}) is rewritten as
\begin{align}\label{MRC_model}
{\tilde{r}_k} = \left( {{\rm{F}}{{\rm{S}}_k}{\rm{ + L}}{{\rm{S}}_k}} \right){x_k} &+ \left( {{\rm{F}}{{\rm{U}}_k}{\rm{ + L}}{{\rm{U}}_k}} \right) {x_k}\notag\\
&+ \left( {{\rm{F}}{{\rm{I}}_k} + {\rm{L}}{{\rm{I}}_k}} \right){x_i} + {\rm{Q}}{{\rm{N}}_k} + {{\rm{N}}_k},
\end{align}
where
\begin{align}
\label{MRC_FS}&{\rm{F}}{{\rm{S}}_k} = \sqrt \rho\sum\nolimits_{m = 1}^{{M_{\rm{f}}}} {\mathsf E}{\left[ {\hat{g}_{{\rm{f}},mk}^H{g_{{\rm{f}},mk}}} \right]} ,\\
\label{MRC_LS}&{\rm{L}}{{\rm{S}}_k} = \alpha \sqrt \rho\sum\nolimits_{m' = 1}^{{M_{\rm{l}}}} {\mathsf E}{\left[ {\hat g_{{\rm{l}},m'k}^H{g_{{\rm{l}},m'k}}} \right]},
\end{align}
\begin{align}
\label{MRC_FI} &{\rm{F}}{{\rm{I}}_k} = \sqrt \rho  \sum\limits_{i = 1,i \ne k}^K {{\hat{\bf{g}}}_{{\rm{f}},k}^H{{\bf{g}}_{{\rm{f}},i}}},~{\rm{L}}{{\rm{I}}_k} = \alpha \sqrt \rho  \sum\limits_{i = 1,i \ne k}^K {{\hat{\bf{g}}}_{{\rm{l}},k}^H{{\bf{g}}_{{\rm{l}},i}}},\\
\label{MRC_FU}&{\rm{F}}{{\rm{U}}_k} = \sqrt \rho  \sum\nolimits_{m = 1}^{{M_{\rm{f}}}} {\left( {\hat g_{{\rm{f}},mk}^H{g_{{\rm{f}},mk}} - {\mathsf E}\left[ {\hat g_{{\rm{f}},mk}^H{g_{{\rm{f}},mk}}} \right]} \right)}  ,\\
\label{MRC_LU}&{\rm{L}}{{\rm{U}}_k} = \alpha \sqrt \rho  \sum\limits_{m' = 1}^{{M_{\rm{l}}}} {\left( {\hat g_{{\rm{l}},m'k}^H{g_{{\rm{l}},m'k}} - {\mathsf E}\left[ {\hat g_{{\rm{l}},m'k}^H{g_{{\rm{l}},m'k}}} \right]} \right)}  ,\\
\label{MRC_N}&{\rm{Q}}{{\rm{N}}_k} = {\hat{\bf{g}}}_{{\rm{l}},k}^H{{\boldsymbol{\omega }}_{\rm{q}}},~~~~{{\rm{N}}_k}= {\hat{\bf{g}}}_{{\rm{f}},k}^H{{\boldsymbol{\omega }}_{\rm{f}}} + \alpha {\hat{\bf{g}}}_{{\rm{l}},k}^H{{\boldsymbol{\omega }}_{\rm{l}}},
\end{align}
and each represents the desired signals contributed from the F-RRHs (FS) and that from the L-RRHs (LS), the interference caused by other users at the F-RRHs (FI) and that at the L-RRHs (LI), the interference due to the channel estimation uncertainty at the F-RRHs (FU) and the L-RRHs (LU), the quantization noise (QN) and the AWGN (N), respectively.

In (\ref{MRC_model}), we treat the sum of the second to the fifth terms in the right hand side as the noise-plus-interference (NPI) power, to which the desired signal is uncorrelated, i.e.,
\begin{equation}\label{Uncorrelated}
{\mathsf E}\left[ {\left( {{\rm{F}}{{\rm{S}}_k}{\rm{ + L}}{{\rm{S}}_k}} \right) \times {\rm{NP}}{{\rm{I}}_k}} \right] = 0.
\end{equation}
Given that uncorrelated Gaussian noise represents the worst case, the achievable SE of the $k$th user for the architecture can be derived by (\ref{achievable_rate}), see top of next page.
\begin{figure*}
\setlength{\abovedisplayskip}{3pt}
\setlength{\belowdisplayskip}{3pt}
\begin{equation}\label{achievable_rate}
{R_k}={\log _2}{\left( {1 + \frac{{{{\left| {{\rm{F}}{{\rm{S}}_k}{\rm{ + L}}{{\rm{S}}_k}} \right|}^2}}}{{{\mathsf E}\left[ {{{\left| {{\rm{F}}{{\rm{U}}_k}{\rm{ + L}}{{\rm{U}}_k}} \right|}^2}} \right] + {\mathsf E}\left[ {{{\left| {{\rm{F}}{{\rm{I}}_k} + {\rm{L}}{{\rm{I}}_k}} \right|}^2}} \right] + {\mathsf E}\left[ {{{\left| {{\rm{Q}}{{\rm{N}}_k}} \right|}^2}} \right] + {\mathsf E}\left[ {{{\left| {{{\rm{N}}_k}} \right|}^2}} \right]}}} \right)}.
\end{equation}
\end{figure*}

\begin{Theorem}\label{Theor_1}
The worst-case achievable uplink SE for the $k$th user in the distributed massive MIMO system with mixed-ADC receivers is lower bounded by (\ref{achievable_Theor_1}), see top of next page.
\begin{figure*}
\begin{equation}\label{achievable_Theor_1}
\setlength{\abovedisplayskip}{3pt}
\setlength{\belowdisplayskip}{3pt}
\resizebox{.94\hsize}{!}
{${R_k} = {\log _2}\left( {1 + \frac{{{{\left( {\sum\limits_{m = 1}^{{M_{\rm{f}}}} {{\beta _{mk}}} } \right)}^2} + 2\alpha \sum\limits_{m = 1}^{{M_{\rm{f}}}} {\sum\limits_{m' = 1}^{{M_{\rm{l}}}} {{\beta _{mk}}{\beta _{m'k}}} }  + {\alpha ^2}{{\left( {\sum\limits_{m' = 1}^{{M_{\rm{l}}}} {{\beta _{m'k}}} } \right)}^2}}}{{\sum\limits_{m = 1}^{{M_{\rm{f}}}} {\sum\limits_{i = 1}^K {{\beta _{mk}}{\beta _{m'i}}} }  + \left( {2\alpha  - {\alpha ^2}} \right)\sum\limits_{m' = 1}^{{M_{\rm{l}}}} {\beta _{m'k}^2}  + \alpha \sum\limits_{m' = 1}^{{M_{\rm{l}}}} {\sum\limits_{i = 1,i \ne k}^K {{\beta _{m'k}}{\beta _{m'i}}} }  + \frac{1}{\rho }\left( {\sum\limits_{m = 1}^{{M_{\rm{f}}}} {{\beta _{mk}}}  + \alpha \sum\limits_{m' = 1}^{{M_{\rm{l}}}} {{\beta _{m'k}}} } \right)}}} \right)$}.
\end{equation}
\end{figure*}
\end{Theorem}

\begin{IEEEproof}
Given that an unbiased channel estimator is employed at each RRH, we have
\begin{align}
\setlength{\abovedisplayskip}{3pt}
\setlength{\belowdisplayskip}{3pt}
\label{E_FS}\left|{\rm{F}}{{\rm{S}}_k}\right|^2  &=  \rho \left( \sum\nolimits_{m = 1}^{{M_\text{f}}} {{\beta _{mk}}}\right)^2 ,\\
\label{E_LS}\left|{\rm{L}}{{\rm{S}}_k}\right|^2  &= {\alpha ^2}\rho {\left( {\sum\nolimits_{m' = 1}^{{M_\text{l}}} {{\beta _{m'k}}} } \right)^2} ,\\
\label{E_FI}{\mathsf E}\left[ {{{\left| {{\rm{F}}{{\rm{I}}_k}} \right|}^2}} \right]&=\rho \sum\nolimits_{m = 1}^{{M_\text{f}}} {\sum\nolimits_{i \ne k}^K {{\beta _{mk}}{\beta _{mi}}} } ,\\
\label{E_LI}{\mathsf E}\left[ {{{\left| {{\rm{L}}{{\rm{I}}_k}} \right|}^2}} \right] &= {\alpha ^2}\rho \sum\nolimits_{m' = 1}^{{M_\text{l}}} {\sum\nolimits_{i \ne k}^K {{\beta _{m'k}}{\beta _{m'i}}} },\\
\label{E_N}{\mathsf E}\left[ {{{\left| {{{\rm{N}}_k}} \right|}^2}} \right] &= \sum\nolimits_{m = 1}^{{M_{\rm{f}}}} {{\beta _{mk}}}  + {\alpha ^2}\sum\nolimits_{m' = 1}^{{M_{\rm{l}}}} {{\beta _{m'k}}},
\end{align}
and ${\mathsf E}\left[ {{ {{\rm{F}}{{\rm{I}}_k}} }} \right]{\mathsf E}\left[ {{ {{\rm{L}}{{\rm{I}}_k}} }} \right]=0$.

Due to the independence among the channel paths to the $M_\text{f}$ F-RRHs, the channel estimation uncertainty at the F-RRHs, ${\mathsf E}[ {{{\left| {{\rm{F}}{{\rm{U}}_k}} \right|}^2}} ]$, can be derived as
\begin{align}
\setlength{\abovedisplayskip}{3pt}
\setlength{\belowdisplayskip}{3pt}
\label{E_FU}{\mathsf E}\left[ {{{\left| {{\rm{F}}{{\rm{U}}_k}} \right|}^2}} \right]& = \rho \sum\limits_{m = 1}^{{M_{\rm{f}}}} {\left( {{\mathsf E}\left[ {{{\left| {\hat g_{{\rm{f}},mk}^H{g_{{\rm{f}},mk}}} \right|}^2}} \right] - {{\mathsf E}^2}\left[ {\hat g_{{\rm{f}},mk}^H{g_{{\rm{f}},mk}}} \right]} \right)} \notag\\
&\mathop  = \limits^a  \rho \sum\nolimits_{m = 1}^{{M_{\rm{f}}}} {\beta _{mk}^2},
\end{align}
where $a$ is obtained since ${\mathsf E}[ {{{| {\hat g_{{\rm{f}},mk}^H{ g_{{\rm{f}},mk}}} |}^2}} ]=2\beta_{mk}^2$. Similarly,
\begin{align}
\label{E_LU}{\mathsf E}\left[ {{{\left| {{\rm{L}}{{\rm{U}}_k}} \right|}^2}} \right]= \alpha^2\rho \sum\nolimits_{m' = 1}^{{M_{\rm{l}}}} {\beta _{m'k}^2}.
\end{align}
We then calculate the power of the quantization noise term $\text{QN}_k$. From \cite{7307134}, we have
\begin{align}\label{E_QN}
{\mathsf E}\left[ {{{\left| {{\rm{Q}}{{\rm{N}}_k}} \right|}^2}} \right]={\mathsf E}\left[ {\left| {{\bf{g}}_{{\rm{l}},k}^H{{\bf{R}}_{{\boldsymbol{\omega} _{\rm{q}}}}}{{\bf{g}}_{{\rm{l}},k}}} \right|} \right],
\end{align}
where ${{\bf{R}}_{{\omega _{\rm{q}}}}}{\rm{ = }}\,\alpha \left( {1 - \alpha } \right){\rm{diag}}( {\rho {{\bf{G}}_{\rm{l}}}\hat{\bf{G}}_{\rm{l}}^H + {\bf{I}}} )$ is the covariance matrix of ${\boldsymbol{\omega}}_\text{q}$. Since the $m'$th diagonal element of ${\rm{diag}} ( {\rho {{\bf{G}}_{\rm{l}}}\hat{\bf{G}}_{\rm{l}}^H + {\bf{I}}} )$ can be expressed as $1 + \rho \sum\nolimits_{i =1}^K {{{| {{\hat g^{H}_{{\rm{l}},m'i}}{g_{{\rm{l}},m'i}}} |}}} $, we can obtain the power quantization noise as (\ref{E_var}).
\begin{figure*}
\begin{align}\label{E_var}
\setlength{\abovedisplayskip}{0pt}
\setlength{\belowdisplayskip}{0pt}
&{\mathsf E}\left[ {\left| {{\bf{\hat g}}_{{\rm{l}},k}^H{\rm{diag}}\left( {\rho {{\bf{G}}_{\rm{l}}}{\bf{\hat G}}_{\rm{l}}^H + {\bf{I}}} \right){{\bf{g}}_{{\rm{l}},k}}} \right|} \right]\notag\\
&= \sum\nolimits_{m' = 1}^{{M_1}} {{\mathsf E}\left[ {\left| {\hat g_{{\rm{l}},m'k}^H{g_{{\rm{l}},m'k}}} \right|} \right]}  + \rho \sum\nolimits_{m' = 1}^{{M_1}} {\sum\nolimits_{i \ne k}^K {{\mathsf E}\left[ {\left| {\hat g_{{\rm{l}},m'k}^H{g_{{\rm{l}},m'k}}} \right|} \right]{\mathsf E}\left[ {\left| {\hat g_{{\rm{l}},m'i}^H{g_{{\rm{l}},m'i}}} \right|} \right]} }+ \rho \sum\nolimits_{m' = 1}^{{M_1}} {{\mathsf E}\left[ {{{\left| {\hat g_{{\rm{l}},m'k}^H{g_{{\rm{l}},m'k}}} \right|}^2}} \right]} \notag\\
&= \sum\nolimits_{m' = 1}^{{M_1}} {{\beta _{m'k}}}  + \rho \sum\nolimits_{m' = 1}^{{M_1}} {\sum\nolimits_{i \ne k}^K {{\beta _{m'k}}{\beta _{m'i}}} }  + 2\rho \sum\nolimits_{m' = 1}^{{M_1}} {\beta _{m'k}^2}.
\end{align}
\hrulefill
\end{figure*}
Substituting (\ref{E_var}) into (\ref{E_QN}), and combining the results from (\ref{E_FS}) to (\ref{E_QN}) obtain the result.
\end{IEEEproof}

Based on \emph{Theorem} 1, the achievable SE is dependent on the transmit SNR, the number of quantization bits $B$, the numbers of F-RRHs and L-RRHs, and the positions of each RRH. The SE obviously increases as the transmit power increases, whereas it has a ceiling due to interference as $\rho \rightarrow \infty$. Also, with ${{\partial {\tt{SINR}}_k} \mathord{\left/
	{\vphantom {{\partial {\tt{SINR}}_k} {\partial \alpha }}} \right.
	\kern-\nulldelimiterspace} {\partial \alpha }} > 0 $ for $\alpha\in\left[0,1\right]$, we conclude that the SE is a monotonically increasing function of the resolution of the receiver and consistent with the trend for the mixed-ADC architecture.
To enhance our understanding of achievable SE, we consider the following two extreme cases.

\begin{Remark}
The SE for the case that all RRHs are equipped with low-resolution ADCs, i.e., $M_\text{f}=0$, gives
\begin{equation}\label{achievable_Ml}
R_k^{{M_{\rm{f}}}{\rm{ = }}0} = {\log _2}{\left( {1 + \frac{{\alpha {{\left( {\sum\nolimits_{m = 1}^M {{\beta _{mk}}} } \right)}^2}}}{{{\tt{J}} + \left( {1 - \alpha } \right)\sum\nolimits_{m = 1}^M {\beta _{mk}^2} }}} \right)},
\end{equation}
where ${\tt{J}}={\sum\limits_{m = 1}^M {\beta _{mk}^2}  + \sum\limits_{m = 1}^M {\sum\limits_{i \ne k}^K {{\beta _{mk}}{\beta _{mi}}} }  + \frac{1}{\rho }\sum\limits_{m = 1}^M {{\beta _{mk}}} }$.
This pure low-resolution ADC architecture can effectively reduce the power consumption and hardware expenditure but suffer performance loss. Note that when $\alpha \rightarrow 1$, (\ref{achievable_Ml}) becomes
\begin{equation}\label{achievable_Mf}
R_k^{{M_{\rm{l}}}{\rm{ = }}0} = {\log _2}\left( {1 + {{{{\left( {\sum\nolimits_{m = 1}^M {{\beta _{mk}}} } \right)}^2}} \mathord{\left/
			{\vphantom {{{{\left( {\sum\nolimits_{m = 1}^M {{\beta _{mk}}} } \right)}^2}} \tt{J}}} \right.
			\kern-\nulldelimiterspace} \tt{J}}} \right),
\end{equation}
which represents the case that all RRHs are F-RRHs. Thus, (\ref{achievable_Mf}) is a closed-form expression given when no quantization noise is introduced by the L-RRHs. This case was considered in most existing works on massive MIMO, despite the high power consumption and hardware cost. 
Moveover, the expression has a similar structure as the result in \cite{123456789}, which was obtained assuming that the channel statistic is estimated via a MMSE detector.
Therefore, the result in (\ref{achievable_Theor_1}) can be regarded as the generalized expression of the achievable SE for distributed MIMO systems with MRC receiver.
%
\end{Remark}

\begin{Remark}
In the centralized massive MIMO case where $\beta_{mk}=\beta_{m'k}=\beta_k$ for $m=1,\ldots,M_\text{f}$, $m'=1,\ldots,M_\text{l}$, the achievable SE is reduced to
\begin{equation}\label{achievable_Co}
{R_k}\!=\!{\log _2}{\left(\!{1\!+ M\frac{{\left( {\kappa  + \alpha \left( {1 - \kappa } \right)} \right){\beta _k}}}{{\frac{1}{\rho } + \sum\limits_{i=1,i \ne k}^K {{\beta _i}}  + \frac{{\kappa  + \left( {1 - \kappa } \right)\left( {2\alpha  - {\alpha ^2}} \right)}}{{\kappa  + \alpha \left( {1 - \kappa } \right)}}{\beta _k}}}} \right)},
\end{equation}
where $\kappa\triangleq {{{M_\text{f}}}  \mathord{\left/
		{\vphantom {{{M_\text{f}}} M}} \right.
		\kern-\nulldelimiterspace} M}$. We note that when all the RRHs are equipped with full-resolution ADC receivers, i.e., $\alpha =1$, the expression is consistent with the result in \cite{Yang}. Hence, summarizing the results of (\ref{achievable_Mf}) and (\ref{achievable_Co}), the achievable SE expression in (\ref{achievable_Theor_1}) can be regarded as the generalized result of the existing works on massive MIMO with MRC receviers.
\end{Remark}

\begin{figure}[ht]
\setlength{\abovecaptionskip}{0cm}
\centering
\includegraphics[width=9.5cm]{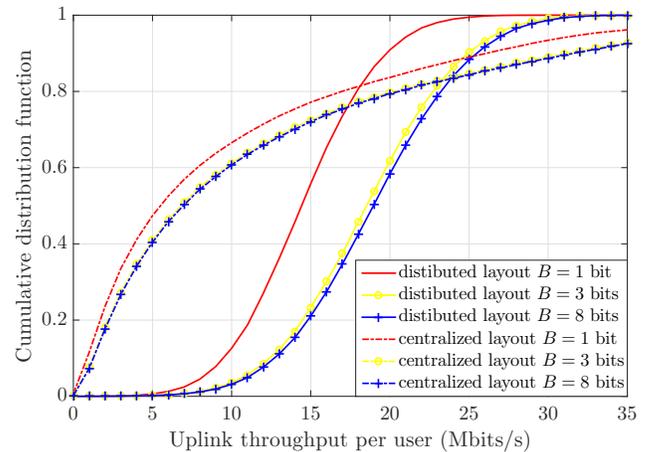}
\caption{The cumulative distribution function (CDF) of the uplink throughput per user, with $M=100$, $K=20$, $\kappa=0.2$ and $\rho=10$\;dB.}
\end{figure}

\section{Numerical Results}


In our simulations, $100$ RRHs and $20$ users were randomly distributed in a $1\,\text{km}^2$ area. We also set a guard zone of 50\,meters for each user, i.e., the distance between any RRH and the user is no less than 50\,m. The decay exponent, shadow-fading standard deviation, and bandwidth were set to $\gamma=3.8$, $\sigma_\text{shad}=8$\,dB, and $W=10$\,MHz, respectively.


We first compare the performance of centralized against distributed massive MIMO setups under mixed-ADC receivers. In Fig.~2, the results for centralized MIMO correspond to those in (\ref{achievable_Co}), in which users are randomly distributed as well. As we can see, the distributed setup significantly outperforms the centralized counterpart, e.g., the 95\%-likely throughput of the architecture with distributed MIMO with 3-bits quantizer is about 11\,Mbits/s, 14\,times more than that of the centralized setup (about 0.8\,Mbits/s). This is mainly because of the spatial diversity and the enhanced signal strength due to reduced distance between the RRHs and the users. More remarkably, the results for L-RRHs with 3-bit ADCs are very close to those for L-RRHs with 8-bit ADCs, indicating that deploying the L-RRHs with a resolution higher than 3\,bits is unnecessary.



\begin{figure}[ht]
\setlength{\abovecaptionskip}{-0cm}
\centering
\includegraphics[width=9.5cm]{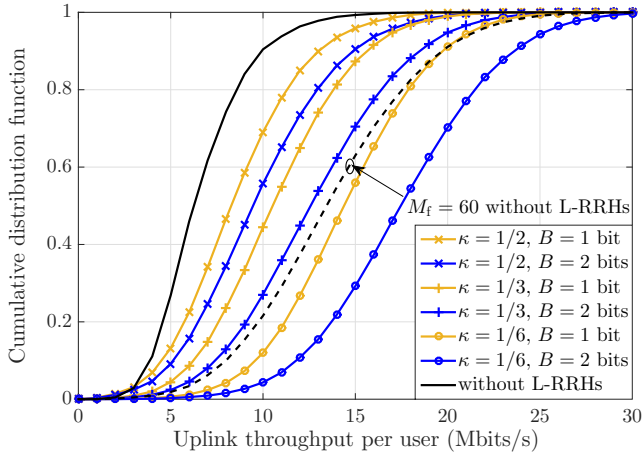}
\caption{The CDF of the uplink throughput per user for distributed massive MIMO with mixed-ADC receivers against with different $\kappa$ and $B$. Results are shown for $M_\text{f}=20$ and $\rho=10$ dB.
}
\end{figure}

Now study the impact of adding L-RRHs on the throughput performance. Fig.~4 shows the CDF of the uplink throughput per user at 10\,dB with 20 F-RRHs in the system. Compared with the result of the pure F-RRHs case, adding a certain number of L-RRHs can significantly improve the system throughput, e.g., the 95\%-likely throughput is increased by about 30\%, 100\%, and 300\% for adding 20, 40, 100 L-RRHs with 2-bit ADCs, respectively. Also, we find that the mixed 120 RRHs architecture with 20 F-RRHs outperforms the pure 60 F-RRHs architecture (denoted by dotted line) about 4\,Mbits/s, indicating that distributed massive MIMO with less full-resolution ADC receivers can achieve superior performance if a large number of L-RRHs can be deployed.

\begin{figure}[ht]
\setlength{\abovecaptionskip}{-0cm}
\centering
\includegraphics[width=9.5cm]{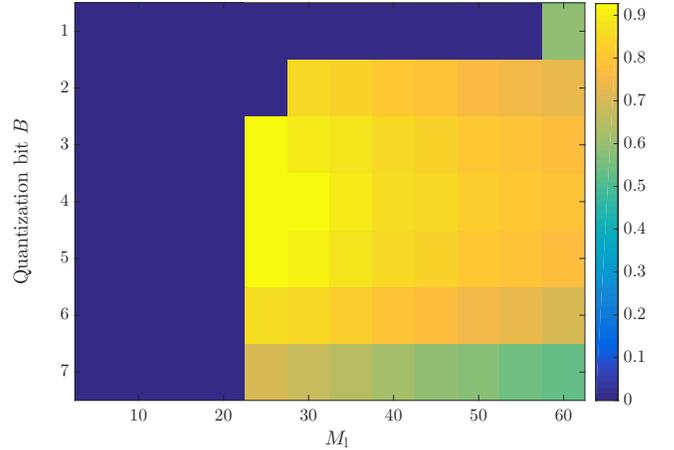}
\caption{The average energy efficiency (in Mbits/Joule) of distributed massive MIMO with mixed-ADC receiver against with quantization bits $B$ and the number of L-RRHs $M_{\rm{l}}$, with $\rho=10$\;dB, $M_{\rm{f}} = 20$, $K = 20$.
}
\end{figure}
Finally, we compare the cost of adding RRHs with that of increasing the resolution bits based on the EE, which is defined as $\eta\,{\rm{ = }}\,{{WR} \mathord{\left/
{\vphantom {{WR} {\left( {{P_{\rm{f}}} + {P_{\rm{l}}}} \right)}}} \right.
\kern-\nulldelimiterspace} {\left( {{P_{\rm{f}}} + {P_{\rm{l}}}} \right)}}$ with
${P_{\left(  \cdot  \right)}} = {M_{\left(  \cdot  \right)}}\left({c_0}{2^{2B}} + {c_1}\right)$, where $c_0\!=\!10^{-4}$\,Watt and $c_1\!=\!2$\,Watt \cite{7307134}. $W$ is the bandwidth set to 10\,MHz, and $R$ is the sum rate.
In Fig. 4, we set a total throughput threshold as $100$\,Mbits/s to guarantee the requirement of users. Adopting the achievable rate in (\ref{achievable_Theor_1}), we observe that the proposed architecture can hardly meet the desired requirement if the number of L-RRHs is less than 20 or the L-RRH is equipped with 1-bits quantizer. We also note the there is a tradeoff between adding L-RRHs and increasing resolution bits of L-RRHs, e.g., adding 60\,L-RRHs with 1-bit quantizer and adding 25\,L-RRHs with 3-bits quantizer can both meet the throughput requirement, whereas we can optimize the L-RRH configuration based on the EE.

\section{Conclusion}
This letter investigated the uplink achievable SE of a multi-user distributed massive MIMO with mixed-ADC receivers. We derived the worst-case achievable SE considering the effect of ADCs. Our results revealed the effects of the system parameters, such as the transmit power and the quantization bits of the low-resolution ADCs, on the achievable SE. The numerical results revealed that the distributed mixed-ADC architecture is an energy-efficient architecture that significantly outperforms the centralized massive MIMO system and can achieve outstanding throughput via deployment of a large number of L-RRHs with appropriate quantization resolution.

\ifCLASSOPTIONcaptionsoff
\fi



%


\end{spacing}



%








\end{document}